\documentclass[aps, singlecolumn, showpacs]{revtex4-2}
\begin{document}
\title{Dirac representation of the group SO(3,2)  and the Landau problem}
\author{S C Tiwari}
\affiliation{Department of Physics, Institute of Science, Banaras Hindu University, Varanasi 221005, and \\ Institute of Natural Philosophy \\
Varanasi India\\ Email address: $vns\_sctiwari@yahoo.com$ \\}
\begin{abstract}
 A systematic study carried out on the infinite degeneracy and the constants of motion in the Landau problem establishes the central extension of the Euclidean group in two dimension as a dynamical symmetry group, and Sp(2,R) as spectrum generating group irrespective of the choice of the gauge. It may be noted that the significance of the Euclidean group was already implicit in the earlier works on the Landau problem; in the present paper the method of group contraction plays an important role. Dirac's remarkable representation of the group SO(3,2) and the isomorphism of this group with Sp(4,R) are re-visited. New insights are gained on the meaning of two-oscillator system in Dirac representation. It is argued that in view of the fact that even the two dimensional isotropic oscillator having SU(2) as dynamical symmetry group does not arise in the Landau problem the relevance or applicability of SO(3,2) group becomes invalid.

\end{abstract}
\maketitle
 
\section{\bf Introduction}

Symmetries and the constants of motion play important role in understanding the kinematics and dynamics of both classical and quantum systems. Degeneracies pose additional challenge in delineating the symmetries of a dynamical system. Nice textbook discussion may be found in \cite{1, 2}. Nonrelativistic quantum theory of electron motion in a constant magnetic field perpendicular to the plane, i. e. the cyclotron motion, has come to be known as Landau problem in the extensive literature on this problem. Infinite degeneracy introduces intricacies as regards to the determination of the dynamical symmetry group for the Landau problem. Since 2D isotropic harmonic oscillator (IHO) emerges in the symmetric gauge for this problem one expects the role of SU(2) symmetry group in some way. Chapter 9 in Goldstein's book presents a good exposition on the half-integral representation of the rotation group SO(3)  in the context of 2D IHO discovered by Jauch and Hill \cite{3}. A thorough analysis by Dulock and McIntosh \cite{4} brings out the complications involved in the Landau problem due to presence of the angular momentum term in addition to the 2D IHO Hamiltonian for the Landau Hamiltonian. In spite of the fact that SU(2) symmetry group appears in this study the constants of motion related with the center of the orbit \cite{5} make the issue difficult. The choice of a gauge and the freedom of gauge transformations make the physical interpretation of the symmetry and the degeneracy in the Landau problem more involved.

Jauch and Hill \cite{3} point out that in special cases degeneracy arises from the invariance of the Hamiltonian under a group of contact transformations leading to dynamical symmetry group in contrast to the invariance under point transformation of the coordinates that give rise to a geometrical group that is a sub-group of the dynamical symmetry group. However, the nature of the transformations that leave the Landau Hamiltonian invariant and the infinite accidental degeneracy have not been settled conclusively for the Landau problem. A systematic study carried out in the present paper establishes that the dynamical symmetry group is the central extension of the Euclidean group E(2) denoted by $\bar{E} (2)$, and the spectrum generating group is Sp(2, R) independent of the choice of the gauge. The former result is already implicit in \cite{4,5}, however we obtain it using the group contraction \cite{6} and the constants of motion derived in \cite{5}. It is remarkable that the Landau Hamiltonian in both symmetric and Landau gauge reduces to 1D HO Hamiltonian for which the Heisenberg algebra can be completed to arrive at SL(2, R) isomorphic to Sp(2, R)  as a spectrum generating group \cite{7}. 

Another important nontrivial issue is related with the question whether wider symmetry group exists for the Landau problem in the spirit of the dynamical non-invariance group or spectrum generating group SO(4,2) for the Hydrogen atom \cite{8}. The Kepler problem in 3D has a new vector constant of motion, namely, the Laplace-Runge-Lenz-Pauli (LRLP) vector \cite{1} and gives rise to compact SO(4) dynamical symmetry group for Hydrogen atom \cite{2}. It would appear surprising that the dynamical symmetry group for the continuum of energy states $(E>0)$ is isomorphic to the non-compact Lorentz group SO(3,1). The group SO(4,2) represents dynamical symmetry group SO(4) as a subgroup for bound states $(E<0)$, SO(3,1) for $(E>0)$ and the Euclidean group in 3D E(3) for $(E=0)$. The group SO(4,2) for Hydrogen atom first appeared in the works of Malkin and Man'ko, and independently of Barut and Kleinert as noted by Kibler \cite{8}. Since some of the generators of SO(4,2) do not commute with the Hamiltonian it is termed as dynamical non-invariance group. Non-commuting generators of this group can be used to generate the whole spectrum for $E<0, E>0, E=0$ hence the terminology the spectrum generating group. Physically motivated nice exposition on SO(4,2) group for the Hydrogen atom can be found in \cite{8}.

In the case of the Landau problem it is not obvious if dynamical non-invariance group makes sense; if one argues that there has to exist such a group one has to find the nature of such a group. Recently, Dereli et al \cite{9} have suggested SO(3,2) group and Dirac's "remarkable representation" \cite{10} for the Landau problem.  Irrespective of the Landau problem two important questions related with Dirac's  paper \cite{10} need a thorough analysis: the meaning of two-oscillator system and the implication of the equivalence of SO(3,2) and Sp(4, R) proved by Dirac towards the end of his paper. It must be realized that Dirac's representation is not restricted to two-oscillator system and two-oscillator system is not limited to 2D IHO. Y. S. Kim has investigated harmonic oscillator systems in various contexts and pointed out that two-oscillator system in Dirac's representation has to be a coupled oscillator system; see \cite{11} for a recent discussion. However, physical examples and conceptual clarity lack on this issue. A detailed discussion and logical arguments presented in the present paper unravel new insights on this fundamental issue. In the context of the Landau problem it turns out that the group Sp(4, R), and SO(3,2) isomorphic to it, are not applicable to account for the dynamical symmetry. In the light of this result some remarks are made on the claim of Dereli et al \cite{9} regarding their motivation based on the duality between 2D Hydrogen atom and the Landau problem vis-a-vis 2D IHO and the assertion that Sp(4, R) is a 'natural symmetry' for the Landau problem. The physical insights gained on two-oscillator system in Dirac's remarkable  representation of SO(3,2), the interesting technical aspects in \cite{9} and the important results on Sp(8, R) underlined in \cite{8} offer prospects for further investigations on fundamental questions in connection with Dirac's two-oscillator system and Landau problem.

The paper is organized as follows. In the next section an overview is presented on 2D IHO, uniform circular motion, and their inter-relationship. Section III reviews the essentials of the Landau problem highlighting the points of present interest. The dynamical symmetry group for the Landau problem is investigated in Section IV. Section V is devoted to gain new insights on the Dirac's representation and the significance of two-oscillator system. Remarks on the applicability of SO(3,2) or Sp(4, R) in the Landau problem are also made in this perspective. Conclusions and outlook constitute the last Section. 

\section{\bf Two-oscillator systems and uniform circular motion}

The cyclotron motion is an ideal uniform circular motion of a point spinless non-radiating electron. Under the application of a uniform magnetic field $B \hat{z}$, the electron motion in the x-y plane is the cyclotron motion and free particle motion along z-axis. In elementary physics course it is taught that the projection of a circulating point on a diameter of the circle executes a simple harmonic motion. A uniform circular motion could be described in terms of the simple harmonic motion of two 1D oscillators at right angles, and having same frequency and amplitude but differing in phase by $90^0$. In a uniform circular motion the magnitude of the tangential velocity is constant, and radial velocity is zero though there is a non-zero radially inward acting centripetal force. Thus, intimate relationship between 2D IHO and cyclotron motion would appear quite natural.

Consider a 2D IHO in cartesian coordinate system; the equations of motion
\begin{equation}
\ddot{x} = - \omega^2 x
\end{equation} 
\begin{equation} 
\ddot{y} = - \omega^2 y
\end{equation}
show that the problem reduces to two independent 1D oscillators. Here $\omega^2 =\frac{k}{m}$ and $k$ is spring force constant. Equations (1)-(2) have simple solutions with four integration constants that are determined by making use of the energy conservation and initial conditions. In polar coordinates $(r,\phi)$ the equations of motion
\begin{equation}
\ddot{r} - r \dot{\phi}^2 = - \omega^2 r
\end{equation}
\begin{equation}
m(r \ddot{\phi} + 2 \dot{r} \dot{\phi})=0
\end{equation}
predict a remarkable result that the angular momentum $ L =m r^2 \dot{\phi}$ is conserved; Eq.(4) is just a statement of this result. Eq.(3) assumes the form
\begin{equation}
\ddot{r} + \omega^2 r = \frac{L^2}{m^2 r^3}
\end{equation}
In cartesian coordinates one can define angular momentum $L=m(x \dot{y} - y \dot{x})$ and calculate it using the solution of the equations (1)-(2). We use the notation $L$ for $L_z$. However, in polar coordinates it directly and explicitly follows from the equations of motion. A 2D IHO problem belongs to the class of the problems of the motion of a particle in central forces: formally it reduces to a 1D problem, and using energy diagrams important qualitative results on the nature of orbits can be obtained \cite{1}.

In this connection the implication of the Bertrand's theorem is worth mentioning: only for Kepler and oscillator problems one has closed orbits for $E<0$ and any value of non-zero angular momentum. The nature of the orbits shows that for Kepler problem one has hyperbolic $(E>0)$, parabolic $(E=0)$, and elliptical $(E<0)$ orbits. For a 2D IHO, if $L$ is non-zero, one has elliptical orbits for all physically admissible energies. For $L=0$, one has straight line motion. From the point of view of the orbit classification, there does exist a partial correspondence between Kepler and oscillator problems.

Integrals or the constants of motion provide deep understanding on the symmetries and degeneracies of a dynamical system. If the generators of infinitesimal canonical transformations leave the Hamiltonian of the dynamical system invariant then these constants of motion form a symmetry group. The angular momentum vector ${\bf L}$ as a generator of rotations in space is a constant of motion, and the symmetry group is SO(3). For the Kepler problem there exists an additional vector constant of motion, namely LRLP denoted by ${\bf A}$. For 2D IHO, the additional constant is a second rank tensor $A_{ij}$. The Poisson brackets for the constants of motion constitute the Lie algebra of the defining symmetry group. For Kepler problem the symmetry group is SO(4) for $E<0$ and SO(3,1) for $E>0$. For a 2D IHO non-zero angular momentum together with $A_{ij}$ tensor show that the symmetry group is SU(2) \cite{1, 3}.

To understand the occrrence of SU(2) rather than SO(3) for the 2D IHO let us consider its Hamiltonian in cartesian coordinates
\begin{equation}
H_o = \frac{1}{2m} (p_x^2 +p_y^2) +\frac{1}{2} m \omega^2(x^2 + y^2)
\end{equation}
Besides $H_o$ there are three constants of motion
\begin{equation}
F_1 = \frac{1}{2} (x p_y - y p_x)
\end{equation}
\begin{equation}
F_2 = \frac{1}{4m \omega} (p_y^2  - p_x^2+ m^2\omega^2(y^2 - x^2))
\end{equation}
\begin{equation}
F_3 = \frac{1}{2 m \omega} (p_x p_y +m^2 \omega^2 x y)
\end{equation}
Since
\begin{equation}
H_o^2 = 4 \omega^2 (F_1^2 + F_2^2 + F_3^2)
\end{equation}
there are only three independent constants of motion. Jauch and Hill \cite{3} remark that $F_3$ "has no obvious physical significance". A clue to the physical interpretation is provided by the orbit characteristics: elliptical orbit needs three independent parameters, namely, the semi-major or minor axis, the eccentricity, and the orientation \cite{1}. Obviously angular momentum is a generator of infinitesimal rotations. Calculating the Poisson brackets it can be shown that $F_i, ~ i =1,2,3$ satisfy the Lie algebra
\begin{equation}
[F_i, F_j] = \epsilon_{ijk} F_k
\end{equation}
In the classical description the origin of SU(2) symmetry group follows from (11), and it is related with the oriented orbits.

McIntosh \cite{12} offers an insightful discussion to show that $F_3$ generates infinitesimal changes in eccentricity preserving the sum of the squares of the semi-axes of the ellipse. A nearly circular orbit is continuously changed by the action of $F_3$ to higher eccentricity ellipses, and finally into a straight line. Continued application of $F_3$ results into reversed traversing to elliptical orbits. A $4\pi$ rotation brings back the orbit to the initial one: two-valuedness or SU(2) symmetry group is because of the oriented orbits.

Instead of phase space variables $(x,y,p_x,p_y)$ having physical dimensions it is convenient to introduce dimensionless variables \cite{9}. In fact, in quantum theory ladder or energy raising-lowering operators or creation-annihilation operators are made dimensionless using $\hbar$. Here we use the variables having uniform dimension: $p_1=(m\omega)^{-\frac{1}{2}} p_x,  p_2=(m\omega)^{-\frac{1}{2}} p_y, q_1=(m\omega)^\frac{1}{2} x, q_2=(m\omega)^\frac{1}{2} y$. In the new variables, the $F_i$ are
\begin{equation}
F_1 = \frac{1}{2} (q_1 p_2 -q_2 p_1)
\end{equation}
\begin{equation}
F_2 = \frac{1}{4} (p_1^2 +q_1^2 -p_2^2 -q_2^2)
\end{equation}
\begin{equation}
F_3 = \frac{1}{2} (p_1 p_2 +q_1 q_2)
\end{equation}
Classically the energies associated with two 1D oscillators are separately constant given by the diagonal elements $A_{11}$ and $A_{22}$ of the tensor $A_{ij}$. However, because of the phase difference, it may be visualized that energy is exchanged between two 1D oscillators during the motion. It becomes more transparent using energy raising-lowering operators in quantum theory \cite{12} constructing a complex quantity
\begin{equation}
F_3 +i F_1 = (p_1 +i q_1)(p_2 - iq_2)
\end{equation}

Thus recognizing that there exist only four constants of motion that do not have explicit time-dependence, and that they are related by Eq.(10) it follows that the Poisson bracket algebra (11) leads to SU(2) dynamical symmetry group for 2D IHO.

Circular cyclotron motion does have a nice correspondence with 2D IHO. The equations of motion of an electron in the constant magnetic field are given by
\begin{equation}
\ddot{x_c} = \omega_c \dot{y_c}
\end{equation}
\begin{equation}
\ddot{y_c} = - \omega_c \dot{x_c}
\end{equation}
Here the cyclotron frequency is $\omega_c = \frac{e B}{m c}$. The force on the electron is calculated from $\frac{e}{c} {\bf v} \times {\bf B}$ for ${\bf B} = B \hat{z}$. A time-dependent rotation of the coordinates
\begin{equation}
x ~ \rightarrow ~ x_c cos \omega_Lt - y_c sin \omega_Lt
\end{equation}
\begin{equation}
y ~ \rightarrow ~ x_c sin \omega_Lt + y_c cos \omega_Lt
\end{equation}
in the oscillator equations (1) -(2) give the cyclotron motion (16)-(17) using the Larmor frequency $2 \omega_L = \omega_c$. To prove it one may use complex representation $z=x+iy$, and the transformation $z \rightarrow z e^{i\omega_Lt}$. Note that the transformation $z \rightarrow z e^{-i\omega_L t}$ would lead to the cyclotron motion in the opposite sense corresponding to ${\bf B} =-B \hat{z}$.

Since 2D IHO is not a coupled oscillator it is worth examining the meaning of a general two-oscillator system. Though 2D oscillator equations (1)-(2) represent a pair of 1D oscillators, these 1D oscillators are not akin to coupled oscillators, for example, in the case of a linear triatomic molecule. A simple example of a coupled oscillator is adding a term $\alpha x y$ to $H_o$ given by the expression (6). Somewhat complicated coupled anharmonic oscillator studied in the literature has the potential energy term 
\begin{equation}
U_o^\prime = \frac{1}{2} (a x^2 +b y^2 +c xy)
\end{equation}

Physically admissible coupled oscillator may have velocity-dependent term in the potential. An illustrative example is that of the time-dependent rotations and seeking invariance of the Hamiltonian (6) under them \cite{13}. The transformation
\begin{equation}
z~ \rightarrow ~z e^{\pm i \alpha(t)}
\end{equation}
together with $\omega \rightarrow \omega \pm \dot{\alpha}$ leaves  the modified Hamiltonian
\begin{equation}
H_o^\prime = H_o\mp \omega L
\end{equation}
invariant. The coupled 2D oscillator system described by expression (22) resembles the Landau Hamiltonian in the symmetric gauge discussed in the next section.

Let us note some aspects on the quantum theory of 2D oscillator. In quantum mechanics the canonical variables $(q_i, p_i)$ turn into operators satisfying Heisenberg commutators. The constants of motion $F_i$ assume the form of the functions of these operators and the Poisson brackets go over to the commutators. Jauch and Hill \cite{3} have underlined the significance of half-integral and integral representations in 2D IHO considering the matrix elements of the operator $F_2$. Since the operator (8) represents half of the difference between the energies of two 1D oscillators the matrix elements of $F_2$ for the quantum states $\Psi_{n_1,n_2}$ of the 2D oscillator would be $(\frac{1}{2} n -n_1)$ where $n_1, n_2$ are the quantum numbers of 1D oscillators and $n=n_1 +n_2$.

The time independent Schroedinger equation for 2D IHO is
\begin{equation}
H_o \Psi =E \psi
\end{equation}
Eq.(23) can be solved either in cartesian or in polar coordinates; however, the nature of the eigenfunctions in two cases is markedly different. In cartesian coordinates the eiegenfunctions are the products of the Hermite polynomials
\begin{equation}
\Psi_n (x,y) =N H_{n_1}(x) H_{n_2} (y)
\end{equation}
where N is a normalization constant. The energy eigenvalues are
\begin{equation}
E_n = (n_1 +n_2 +1) \hbar \omega
\end{equation}
The degree of degeneracy is n+1. Solving in polar coordinates, the $\phi$ dependence is just $e^{iM\phi}, M=0,\pm1,\pm2...$~. The energy eigenvalues are
\begin{equation}
E_n = (2 n_r +|M|+1) \hbar \omega
\end{equation}
The degree of degeneracy is n+1 where $n= 2 n_r +|M|$, and $n_r$ is radial quantum number. The eigenfunctions are obtained in terms of the Laguerre polynomials 
\begin{equation}
\Psi_n(r,\phi) = N^\prime e^{iM \phi} L^M_{n_r}(r)
\end{equation}

The crucial difference between (24) and (27) is that (27) is an eigenfunction of the angular momentum operator
\begin{equation}
L \Psi_n(r,\phi)=-i\hbar \frac{\partial}{\partial \phi} \Psi_n(r,\phi) = M\hbar \Psi_n(r,\phi)
\end{equation}
whereas it is only for the ground state n=0 that (24) is an eigenstate of angular momentum with zero eigenvalue. One can construct angular momentum eigenstates using linear superposition of the degenerate states in cartesian coordinate system using (24). Is this strange feature a peculiarity of quantum description? Recalling the discussion on the classical treatment of 2D IHO, it becomes clear that this feature is a manifestation of the difference between the coordinate systems: explicit angular momentum conservation arises in the equation of motion in polar coordinates $(r,\phi)$ in Eq.(4) that is absent in Eqs. (1) and (2) in cartesian coordinates.

Almost everything discussed in this section can be found, though, in pieces scattered in the textbooks/literature. A coherent picture combining the pieces together has led to new insights on the correspondence between 2D IHO and uniform circular motion, the intriguing role of angular momentum, and the significance of SU(2) group as a dynamical symmetry group in both classical and quantum description of 2D IHO.

\section{\bf Aspects of the Landau problem}

A uniform magnetic field in z-direction can be obtained from a vector potential in a symmetric gauge
\begin{equation}
{\bf A} =\frac{B}{2} (-y \hat{x} +x \hat{y} +0\hat{z})
\end{equation}
or Landau gauge
\begin{equation}
{\bf A} = B (0\hat{x} +x\hat{y} +0\hat{z} )
\end{equation}
The Hamiltonian for the electron motion in the x-y plane for the symmetric gauge is 
\begin{equation}
H_L = \frac{1}{2m} ({\bf p} +\frac{e{\bf A}}{c})^2 = H_o +\omega_L L
\end{equation}
Here $H_o$ is 2D IHO Hamiltonian (6) replacing $\omega$ by $\omega_L$. The Schroedinger equation for the Landau problem is
\begin{equation}
H_L \Psi = E\psi
\end{equation}
Energy eigenfunctions and eigenvalues for the Landau problem can be found in textbooks/review articles. The motivation in the present paper is to unravel the distinct characteristics of degeneracy and symmetries, therefore, the differences and the similarities of the Landau problem with 2D IHO and 2D Hydrogen atom \cite{3,4,5,9} are highlighted.

{\bf [A1]: Energy eigenvalues and quantum numbers}

The Hamiltonian $H_L$ differs from 2D oscillator Hamiltonian by a term proportional to the angular momentum. Since $L$ commutes with $H_L$, it is an integral of motion. The eigenfunctions of $H_L$ are those that are simultaneously the eigenfunctions of $H_o$ and $L$. The energy eigenvalues are just the sum of (26) and $M\hbar \omega_L$
\begin{equation}
E_{n_r,M} =(2n_r+|M| +1 +M) \hbar \omega_L
\end{equation}
For $M$ negative the energy eigenvalues (33) are
\begin{equation}
E_{n_r} = (n_r +\frac{1}{2})\hbar \omega_c
\end{equation}
One finds that the Landau problem has infinite degeneracy in contrast to 2D IHO that has finite degeneracy n+1.

As regards to the correspondence of the Landau problem with 2D Hydrogen atom it has to be realized that the Levi-Civita transformation shows only partial equivalence even with 2D IHO \cite{14}. In the case of the Landau problem the energy eigenvalues given by the expression (34) are equivalent to 1D oscillator that would make the correspondence with 2D hydrogen atom questionable.

{\bf [A2]: Center of the orbit, degeneracy and conjugate variables}

New constants of motion $(x_0,y_0)$ interpreted as the coordinates of the center of the orbit \cite{5} show quantum behavior in the sense that they commute with $H_L$ but do not commute with each other having the nonvanishing commutator
\begin{equation}
[x_0, y_0] = i \lambda^2
\end{equation}
where the length $\lambda =(\frac{\hbar c}{eB})^\frac{1}{2}$. Infinite degeneracy is explained using the fact that the eigenvalues of $x_0$ or $y_0$ have the infinity of real numbers, and the eigenfunctions of $H_L$ include an infinte manifold of the eigenfunctions of $x_0$ or $y_0$. Non-commuting pair $x_0,y_0$ implies that the center of the orbit is indeterminate, and $\Delta x_0 ~\Delta y_0 \geq \frac{\lambda^2}{2}$. Johnson and Lippmann \cite{5} calculate eigenfunctions of $x_0$ and show that considering the eigenvalues of $r_0^2$
\begin{equation}
r^2_0 =x^2_0 +y^2_0
\end{equation}
the energy eigenvalues of the Landau problem correspond to 1D oscillator. 

We suggest an alternative interpretation of $(x_0,y_0)$ using their definition
\begin{equation}
x_0 =\frac{x}{2} - \frac{p_y}{m \omega_c}
\end{equation}
\begin{equation}
y_0 =\frac{y}{2} + \frac{p_x}{m \omega_c}
\end{equation}
and re-writing the commutator (35) in the form of canonically conjugate coordinate and momentum variables $(X,P)$
\begin{equation}
X=\frac{x}{2} + \frac{p_y}{m \omega_c}
\end{equation}
\begin{equation}
P = p_x - \frac{m\omega_c y}{2}
\end{equation}
The commutator is
\begin{equation}
[X, P] =i \hbar
\end{equation}
and the Hamiltonian (31) in terms of $(X, P)$ reads
\begin{equation}
H_L = \frac{P^2}{2m} + \frac{1}{2} m \omega_c^2 X^2
\end{equation}
Expression (42) is a 1D oscillator Hamiltonian, and the 2D oscillator problem is reduced to the 1D problem along with $\omega_L \rightarrow \omega_c$. This result is in exact correspondence with that given in \cite{5} and also with the energy eigenvalue expression (34).

To understand infinite degeneracy let us note that another pair $(X^\prime, P^\prime)$ given by
\begin{equation}
X^\prime =\frac{x}{2} - \frac{p_y}{m \omega_c}
\end{equation}
\begin{equation}
P^\prime = p_x + \frac{m\omega_c y}{2}
\end{equation}
also represents canonically conjugate variables, and commutes with $(X, P)$. However, the 1D oscillator Hamiltonian 
\begin{equation}
H_L^\prime = \frac{{P^\prime} ^2}{2m}+\frac{1}{2} m \omega_c^2 {X^\prime}^2
\end{equation}
represents the Landau Hamiltonian in 2D with reversed direction of the orbit
\begin{equation}
H^\prime_L =H_o -\omega_L  L
\end{equation}
The energy eigenvalues for (46) would become
\begin{equation}
E^\prime_{n_r,M} = (2n_r +|M|+1 -M) \hbar \omega_L
\end{equation}
and reduce to (34) for the positive values of M. The infinite degeneracy arises because $(X^\prime, P^\prime)$ do not appear in $H_L$. It may be pointed out that one can re-write directly $H_L$ or $H_L^\prime$ as some of the squared terms (42) or (45) without following the route via the conjugate variables $(x_0, y_0)$ introduced in \cite{5}.

{\bf [A3]: Gauge invariance in the Landau problem}

A footnote in \cite{5} and one of the sections of \cite{4} make essential points regarding the role of gauge transformation and gauge invariance in the Landau problem. The magnetic field ${\bf B} ={\bf \nabla} \times {\bf A}$ implies that ${\bf B}$ remains unaltered if ${\bf A} \rightarrow {\bf A} + {\bf \nabla}\chi$. It is easily verified that the symmetric gauge (29) goes to the Landau gauge choosing
\begin{equation}
\chi=\frac{B}{2} xy
\end{equation}
The Hamiltonian (31) in the Landau gauge is given by 
\begin{equation}
H_L = \frac{1}{2m} (p_y +\frac{eB}{c} x)^2 +\frac{p_x^2}{2m}
\end{equation}
In this case $p_y$ commutes with $H_L$, and one can solve the Schroedinger equation assuming
\begin{equation}
\Psi(x,y) =e^{ik_y y} \Psi (x)
\end{equation}
The problem reduces to the 1D oscillator problem with
\begin{equation}
H_L =\frac{p_x^2}{2m} = \frac{1}{2} m \omega_c^2(x+ \frac{\hbar c}{eB} k_y)^2
\end{equation}
Formally one has equivalent description if $\chi = -\frac{B}{2} xy$ instead of (48).

In both cases there is a translational symmetry in contrast to the rotational symmetry for the symmetric gauge. In all the three gauges the energy eigenvalues correspond to 1D oscillator with angular frequency $\omega_c$. Dulock and McIntosh \cite{4} show that in the symmetric gauge the center of the orbit is translated using a gauge transformation, and it is rotated using the rotated coordinates as well as the rotated vector potential. Now, the Hamiltonian $H_o^\prime$ given by expression (22) is indistinguishable from the Landau Hamiltonian (31) in the symmetric gauge. Therefore, the symmetries and degeneracies for the two problems are common irrespective of the issues related with gauge invariance. The translational symmetry in this case would correspond to that of the location of the center of the orbit in x-y plane. The formal manipulation from (37) to (41) also holds for the coupled 2D oscillator described by $H_o^\prime$. The distinct feature of the Landau problem where gauge transformation becomes important is in the alternative possibility of the description in the Landau gauge or the alternative with $\chi =-\frac{B}{2} xy$.

\section{\bf Symmetry group in the Landau problem}

The accidental degeneracy in 2D IHO found explanation in terms of the dynamical symmetry group SU(2). Is the group SU(2) also a symmetry group for the Landau problem? In the last two sections new insights have been gained showing that in spite of the significance of 2D IHO in the context of the cyclotron motion and the Landau problem there exist fundamental differences not sufficiently recognized in the literature. The most crucial is that of infinite degeneracy in the Landau problem irrespective of the choice of gauge in contrast to the oscillator problem that has finite degree of degeneracy. It is for this reason that SU(2) symmetry group cannot explain the degenerate Landau problem. Yet, SU(2) group does appear in the discussions on the Landau problem. To delineate the role of SU(2) group and establish the right symmetry group become the important issues. 

The role of translational symmetry in explaining the infinite degeneracy as well as the connection between the Euclidean group E(2) and infinite degeneracy have also been discussed in the past. The emergence of an inner symmetry related with the center of the orbit and outer symmetry in the x-y plane as regards to the location of the orbit is pointed out in \cite{4} following the work of Johnson and Lippmann \cite{5}.

In order to investigate the question of dynamical symmetry group it is worth to examine the so called Zeeman effect for a harmonic oscillator studied by Dulock and McIntosh \cite{4}.  An attractive harmonic force is added to the cyclotron motion, and this leads to the modified Landau Hamiltonian for this problem
\begin{equation}
H_Z = H_L +\frac{1}{2} m \omega^2 (x^2 + y^2)
\end {equation}
Here $\omega$ is the angular frequency of the oscillator. In the classical problem matrix representation of (52) as an operator under Poisson brackets is considered, and eigenvalues and eigenfunctions $(u, v)$ of this matrix operator are calculated assuming the basis of space of coordinates and momenta. The constants of motion are the products of the eigenfunctions the sum of whose eigenvalues are zero: $uu^*, vv^*, u^{*R}v, u^Rv^*$. Here R is given by $R\lambda_1 =\lambda_2$, and
\begin{equation}
\lambda_1 = (\omega^2 + \omega_L^2)^\frac{1}{2} + \omega_L
\end{equation}
\begin{equation}
\lambda_2 = (\omega^2 + \omega_L^2)^\frac{1}{2} - \omega_L
\end{equation}
A linear combination of the constants of motion is constructed to define four quantities $(H_Z, K, L, D)$. Here $H_Z$ is the Hamiltonian. Poisson bracket algebra shows that $(K, L, D)$ commute with the Hamiltonian, and satisfy the SU(2) Lie algebra. 

Unfortunately, the discussion given in \cite{4} gives unintended misleading impression as if SU(2) symmetry group is important for the cyclotron/Landau problem. Two limiting cases for the constants of motion corresponding to 2D oscillator and pure cyclotron motion seem to clarify the situation; however, the way general case for $H_Z$ is presented is not satisfactory: the presence of the harmonic potential in $H_Z$ removes infinite degeneracy and the constants of motion are just the ones generalized for 2D oscillator akin to $(F_1, F_2, F_3)$ defined by expressions (7)-(9). It is, therefore, natural to expect that $(D, K, L)$ satisfy SU(2) Lie algebra. In fact, it is straightforward to calculate the energy eigenvalues for $H_Z$ in exactly the way one derives (33)
\begin{equation}
E^Z_{n_r,M} = (2n_r +|M| +1) \hbar (\omega^2_L +\omega^2)^\frac{1}{2} +M \hbar \omega_L
\end{equation}
The difference between the frequencies multiplying $|M|$ and M in (55) shows that infinite degeneracy is removed, and in spite of the presence of the coupled angular momentum term $\omega_L L$ in $H_Z$ the dynamical symmetry group has to be that of 2D IHO, i. e. SU(2). 

The limiting cases discussed in \cite{4} also show that SU(2) is not a symmetry group for the cyclotron motion. Though Wigner's group contraction is noted in \cite{4} it is not used to get the symmetry group for the Landau problem. Here we show that the group contraction method makes the derivation of dynamical symmetry group for the Landau problem quite insightful. The motivation for group contraction could be understood in a simple manner. A physical theory may correspond to a limiting case of a more generalized theory: relativistic mechanics goes over to nonrelativistic Newtonian mechanics in the limit of the velocity of light tending to infinity. Motivated by this well-known theory reduction in physics Inonu and Wigner \cite{6} introduce the idea of group contraction: underlying symmetry group of one theory may correspond to the limiting case of the symmetry group of the other theory. Group contraction in Wigner's method actually deals with the contraction of the Lie algebras of the groups with respect to any of the continuous subgroups: generators of the algebra and their commutation relations. The basis elements of the linear space defining the Lie algebra can be changed by a non-singular transformation then the transformed Lie algebra describes the isomorphic algebra to the original one. If a singular transformation is made and the transformed generators fulfil the conditions of a Lie algebra then one arrives at a new group.

A simple example of the contraction of the inhomogeneous Lorentz group in 1+1 D space-time to the inhomogeous Galilie group in 1D space taking the limit $c \rightarrow \infty$ explains the method of contraction and the significance of the singular transformation carried out in a way that respects the representation of the contracted group \cite{6}. Since the transformation of basis elements is made in a sequence depending on some parameters $\epsilon_i \rightarrow 0$, the contraction may lead to different groups depending on the sequence. A transformation is singular if all the generators depending on $\epsilon$ tend to zero either in the same way as $\epsilon \rightarrow 0$ or some of them vanish faster than $\epsilon \rightarrow 0$. The example of SU(2) group contraction is interesting and relevant here. Let us consider the Lie algebra (11)
\begin{equation}
[F_1, F_2] =F_3, ~ [F_2, F_3] = F_1, ~ [F_3, F_1] = F_2
\end{equation}
If $F_i \rightarrow F_i^c = \epsilon F_i, i=1,2$ and $F_3 \rightarrow F_3^c = F_3$ then the algebra of the contraction group is that of E(2) letting the limit $\epsilon \rightarrow 0$
\begin{equation}
[F_1^c, F_2^c] =0, ~ [F_2^c, F_3^c] = F_1^c, ~ [F_3^c, F_1^c] = F_2^c
\end{equation}
For the second kind of sequence $F_i \rightarrow F_i^c = \epsilon F_i, i=1,2$ and $ F_3 \rightarrow F_3^c= \epsilon^2 F_3$, in the limit $\epsilon \rightarrow 0$ we have the following contraction algebra
\begin{equation}
[F_1^c, F_2^c] = F_3^c, ~ [F_2^c, F_3^c] = [F_3^c, F_1^c]=0
\end{equation}
It may be pointed out that the Heisenberg commutators $[x_i, p_j] = i\hbar \delta_{ij}$ define the Heisenberg algebra; the contraction algebra (58) is nothing but the Heisenberg algebra.

The limiting cases in the Zeeman effect problem described by $H_Z$ can be treated using the method of group contraction. The limiting case $\omega_L \rightarrow 0$ corresponds to a trivial case of a nonsingular transformation leading to $SU(2) \rightarrow SU(2)$ symmetry group. The second limiting case $\omega \rightarrow 0$ represents a singular transformation in which $R \rightarrow 0$. The new algebra discussed in \cite{4} in this limiting case obtained from the SU(2) Lie algebra for the Hamiltonian $H_Z$ is reproduced below
\begin{equation}
[(v^* +v), (v^* - v)] =- 8 \omega_L i
\end{equation}
\begin{equation}
i[(uu^*)^\frac{1}{2} (v^* -v), uu^*]=0
\end{equation}
\begin{equation}
[uu^*, (uu^*)^\frac{1}{2} (v^* +v)] =0
\end{equation}
where
\begin{equation}
u = \omega_L m^\frac{1}{2} (x+iy) +i m^{-\frac{1}{2}} (p_x +i p_y)
\end{equation}
\begin{equation}
v = \omega_L m^\frac{1}{2} (x-iy) +i m^{-\frac{1}{2}} (p_x -i p_y)
\end{equation}
In the contraction method we assume $\epsilon = R^\frac{1}{2}$ and make the transformation of the SU(2) generators $K \rightarrow \epsilon K, ~L \rightarrow \epsilon L,~ D \rightarrow \epsilon^2 D$. Now, taking the limit $\epsilon \rightarrow 0$ leads to the Heisenberg algebra (59)-(61). It is easily verified that the constants of motion $(v+v^*)$ and $(v -v^*)$ are essentially $(x_0, y_0)$ defined by expressions (37)-(38) and denoted as $(S, Q)$ in \cite{4}.

To find the proper constants of motion in a problem having accidental degeneracy is not a trivial or easy task \cite{3}.The infinite degeneracy in the Landau problem introduces more complications. In the symmetric gauge, the non-commuting constants of motion $(x_0, y_0)$ would serve the purpose if one more proper constant of motion could be found; the angular momentum seems a good choice to construct the Lie algebra of the dynamical symmetry group. The commutator (35) taking the variable $y_0$ in the dimension of momentum $p=m \omega_c y_0$ becomes
\begin{equation}
[x_0, p]= i \hbar
\end{equation}
In the notation of dimensionless variables $x_0 \rightarrow q_0, ~ p \rightarrow p_0$ the Heisenberg commutators of $q_0, p_0$ with L can be easily calculated to give
\begin{equation}
[q_0, L ] =-i \hbar p_0
\end{equation}
\begin{equation}
[p_0, L] = i \hbar q_0
\end {equation}
The Lie algebra defined by (64)-(66) is peculiar for two reasons: (i) the algebra has first order quantities $(q_0, p_0)$ as well as the quadratic one $L$, therefore, it differs from the algebra of SU(2) group, and (ii) it has Heisenberg algebra (64) thereby differing from the algebra of the Euclidean group (57). In fact, the Lie algebra (64)-(66) defines the central extension of the group E(2) denoted by $\bar{E}(2)$ in the literature. Note that the algebra defined by $(H, D, S)$ for the cyclotron motion \cite{4} is same as the algebra of the group $\bar{E}(2)$.

The most remarkable characteristic of the Landau problem is that irrespective of the choice of the gauge it reduces to 1D HO Hamiltonian. A gauge-independent symmetry group could be envisaged for 1D oscillator. Let us consider the Hamiltonian (42), and the Heisenberg algebra (41) for the variables $(X, P)$. The algebra (41) can be completed to "strictly quadratic" algebra \cite{7} defining
\begin{equation}
I_1 = \frac{1}{4} (XP+PX)
\end{equation}
\begin{equation}
I_2 = \frac{1}{4} (X^2 - P^2)
\end{equation}
\begin{equation}
I_3 =-\frac{1}{4} X^2 + P^2)
\end{equation}
we get the following commutators in the units $\hbar=1$
\begin{equation}
[I_1, I_2] =i I_3, ~ [I_2, I_3] =-i I_1, ~ [I_3, I_1] = -i I_2
\end{equation}
The algebra (70) is the Lie algebra of the group SL(2, R)=Sp(2, R) =SO(2,1). The group SL(2, R) is the spectrum generating group; the generators $I_j, j=1,2,3$ do not commute with the Hamiltonian (42). We get new result using group contraction of SU(2): the contraction process in two different ways leads to Heisenberg algebra whose completion determines the spectrum generating group Sp(2, R) isomorphic to SL(2, R), and to the algebra of the group E(2) whose central extension leads to the group $\bar{E}(2)$ that determines the dynamical symmetry of the Landau problem.
 
Physically intuitive picture for the infinite degeneracy in tha Landau problem is that the eigenvalues of $x_0$ and $y_0$ constitute the infinity of the continuum in $R^2$ and the energy eigen value is independent of the location of the center of the orbit in the x-y plane. In the Landau gauge the Hamiltonian is invariant under the space translation along the y-axia, and the energy eigenvalue is independent of the momentum along y-direction. Recalling the mapping of 1D harmonic oscillator motion on the uniform circular motion it follows that the choice of the other gauge in which the Hamiltonian is invariant under space translation along x-axis and the energy eigenvalue is independent of the momentum along x-direction has the complementary role for the complete mapping of the pair of 1D oscillators.

\section{\bf Dirac representation of SO(3,2) group}

In particle physics the spectrum-generating group aproach using the group SO(3,2) has been extensively studied in the literature. Ehrman \cite{15} has discussed the universal covering group of SO(3,2) and the invariants formed from the polynomials of the generators for irreducible representations. It is important to note that both the groups SO(4,1) and SO(3,2) can be contracted to the Poincare group using Wigner's method \cite{6}, however, it is only for SO(3,2) group that irreducible unitary representations exist possessing positive-definite or negative-definite energies. Since SO(3,2) group is a noncompact group the quadratic Casimir operator and fourth-order invariant \cite{15} related with mass and spin of a particle respectively may not be sufficient for irreducible representations. Dirac's remarkable representation \cite{10} resolves this issue using $m_{\mu 5}$, four out of the ten generators $m_{ab}$, defined in Dirac's paper. Majorana representation is Dirac's remarkable representation in which the eigenvalues of $m_{\mu 5}$ play the role of additional invariant to label the irreducible representations. In a recent paper \cite{9} the use of SO(3,2) and Dirac's remarkable representation of this group has been proposed for the dynamical symmetry of the Landau problem. This suggestion is interesting and provocative to re-visit Dirac's representation to understand fundamental questions related with the symplectic group Sp(4,R) and two-oscillator system appearing in \cite{10}.  This enables us to make a critical appraisal on the suggested application of the group SO(3,2) in the Landau problem \cite{9}. In this section we first discuss the fundamental questions related with the Dirac representation in generality, and then analyze the Landau problem in this perspective.

The group of rotations that leaves the quadratic form
\begin{equation}
g_{ab} x^a x^b = x_1^2 +x_2^2 + x_3^2 - x_4^2 - x_5^2
\end{equation}
invariant defines the 3+2 de Sitter group SO(3,2). Here the indices $a, b = 1,2,3,4,5$ and $\mu, \nu =1,2,3,4$. The antisymmetric generators of the Lie algebra of the group $m_{ab}=-m_{ba}$ satisfy the commutation relations
\begin{equation}
[m_{ab}, m_{cd}] =0
\end{equation}
if $a,b,c,d$ are all different, and
\begin{equation}
[m_{ab}, m_{ac}] =m_{bc} ~ ~ a=4,5
\end{equation}
\begin{equation}
[m_{ab}, m_{ac}] = -m_{bc} ~~ a=1,2,3
\end{equation}
where $a,b,c$ are all different. In Dirac's notation, $i m_{ab}$ for all $a,b$ have real eigenvalues for unitary representations. In the "new representation" based on complex variables Dirac introduces a set of real variables $(q_1, p_1, q_2, p_2)$ to get the expressions for all $i m^\prime_{ab}$, and remarks that they satisfy the commutators $[q_i, p_j] =i \delta_{ij}, ~i,j =1,2$ similar to those of canonically conjugate variables in quantum mechanics. It is noted that $i m^\prime_{45}$ 
\begin{equation}
i m^\prime_{45} = \frac{1}{4} (q_1^2 + p_1^2 +q_2^2+p_2^2)
\end{equation}
resembles the half of the sum of the energy of two 1D oscillators.  Besides these observations there is no mention of two-oscillator system in \cite{10}. At the end of the paper following a remark by R. Jost Dirac considers the group Sp(4, R) and proves its equivalence with the group SO(3, 2). Since the symplectic group Sp(4, R) is a dynamical group for 2D oscillator it is tempting to assert that Dirac's remarkable representation deals with two-oscillator system; however, it has to be carefully analyzed in what sense Sp(4, R) arises for 2D oscillator.

Following \cite{10} we have four of the ten generators to be cyclic
\begin{equation}
i m^\prime_{12} = \frac{1}{2} (q_1 p_2 - q_2 p_1)
\end{equation}
\begin{equation}
i m^\prime_{23} = \frac{1}{4} (q_1^2 + p_1^2 - q_2^2  -p_2^2)
\end{equation}
\begin{equation}
i m^\prime_{31} = -\frac{1}{2}(q_1 q_2 + p_1 p_2)
\end{equation}
and $i m^\prime_{45}$ given by (75). Comparing (76)-(78) with (12)-(14) it becomes clear that $(i m^\prime_{12}, im^\prime_{23}, i m^\prime_{13})$ satisfy SU(2) Lie algebra (11) of 2D IHO. The interesting point in Dirac's representation is that the angular momentum operator $i m^\prime_{12}$ has half-odd eigenvalues for odd wavefunctions, and the energy eigenvalues for $i m^\prime_{45}$ have integral positive values. The set of the operators $(i m^\prime_{12}, im^\prime_{23}, i m^\prime_{13})$ satisfies the sub-algebra isomorphic to the SU(2) algebra whereas the set of the generators $(i m^\prime_{45}, i m^\prime_{43}. i m^\prime_{35})$ satisfies a sub-algebra isomorphic to that of the concompact group SO(2,1) equivalent to Sp(2, R). It is clear that SU(2) sub-algebra satisfied by only a sub-set of the generators of the group SO(3,2) in Dirac representation corresponds to a 2D IHO, and two-oscillator system is not 2D IHO. 

One of the best, and perhaps earliest example of a two-oscillator system in Dirac's remarkable representation of the 3+2 de Sitter group is the one used by Dirac himself \cite{16,17}. Positive energy wave equation, Majorana equation in a different representation, was derived using six generators $s_{\mu\nu}$ that are spin operators arising in the Lorentz group, and formally additional quantities $s_{\mu 5}$ were constructed \cite{16}. The ten operators defined in this way are similar to the generators $m_{ab}$ of the SO(3,2) group \cite{10}. Here the two-oscillator system represents the internal degrees of freedom. In the second paper a physical interpretation of $s_{\mu 5}$ is sought in terms of a pulsating spherical shell: this implies that the two-oscillator system necessitates an extended structure.  

Non-compact symplectic group Sp(2N, R) $N \geq 1$ may be viewed as a dynamical non-invariance group for isotropic oscillator in N dimension, and the dynamical symmetry group is the maximal compact sub-group SU(N) \cite{8}. The important point to note is that for IHO with attractive potential only SU(N) appears as a dynamical symmetry group. Therefore, Dirac's proof that Sp(4, R) is equivalent to SO(3,2) has to be viewed in this sense of a two-oscillator system: both attractive and repulsive potentials are needed.

The perspective put forward here for Dirac's remarkable representation of the group SO(3,2) \cite{10} offers new insights on the question of its applicability to the Landau problem. Since the relevant symmetry groups for the Landau problem established in the preceding section are $\bar{E}^2$ and Sp(2, R) it follows that the group SO(3,2) or its contractions have no relevance for the Landau problem. How do we understand the claim made by the authors \cite{9} contrary to this logical conclusion? A thorough examination of some important issues in this context becomes necessary including two crucial points made by Dereli et al \cite{9}: the duality between 2D Hydrogen atom and the Landau problem, and the claim that the group Sp(4, R) is a natural symmetry for the Landau problem.

{\bf B1: Hydrogen atom - oscillator correspondence}

The remarkable correspondence between Hydrogen atom and IHO noted since long has been extensively discussed in the literature. A lucid presentation can be found in \cite{8}. Physically obvious symmetry to explain the observed 2l+1 - fold degeneracy in Hydrogen atom for the quantum number m that can vary from -l to +l in integral steps is the rotational symmetry of the Coulomb potential. The orbital angular momentum operators ${\bf L}$ define the Lie algebra of the group SO(3). The accidental degeneracy is due to l-degeneracy: for each value of the principal quantum number n, l can vary from 0 to n-1. The LRLP vector ${\bf A}$ together with ${\bf L}$ define the Lie algebra of the group SO(4). The commuting generators $(H, L^2, L_z)$ define SO(3) Lie algebra, and this group is termed a geometrical symmetry group. The set of commuting operators $(H, L^2, L_z, A^2, A_z)$ define the dynamical symmetry group SO(4) to account for the accidental degeneracy. Hydrogen atom has discrete energy spectrum $(E< 0)$. IHO potential also has rotational symmetry, and the dynamical symmetry group for 3D IHO is SU(3) \cite{2, 8}.

In two dimensional systems, the accidental degeneracy for the Hydrogen atom is explained by the dynamical symmetry group SO(3), and that of IHO by the group SU(2) \cite{3}. Levi-Civita transformation shows the duality between 2D Hydrogen atom and 2D IHO. However, a careful study on the generators of the symmetry groups in two cases shows that the equivalence is only partial \cite{14}. Since SU(2) is a covering group of SO(3) this partial equivalence is already present implicitly in \cite{3}.

We have shown here that the Landau problem is definitely different than 2D IHO, therefore, even the partial correspondence between 2D Hydrogen atom and 2D IHO has no significance in connection with the dynamical symmetry of the Landau problem.

{\bf B2: Dynamical non-invariance group}

The wider symmetry group SO(4,2) can explain the whole energy spectrum of the Coulomb problem in 3D for $E<0,~E>0,~E=0$ with the corresponding dynamical symmetry groups SO(4), SO(3,1) and E(3) respectively \cite{8}. Note that SO(3,1), E(3) and SO(4,2) are non-compact groups whereas SO(3) and SO(4) are compact groups. The group SO(4,2) is a dynamical non-invariance group as some of its generators do not commute with the Hamiltonian. However, the action of the non-commuting generators on an eigenstate leads to another state, therefore, one may term SO(4,2) as spectrum generating group. Unitary irreducible representations of SO(4,2) could also be realized in the non-compact sub-groups SO(4,1) and SO(3,2). Thus, the group SO(3,2) is also a dynamical non-invariance group  for the Coulomb problem in 3D.

{\bf B3: Symplectic group and isotropic harmonic oscillator}

The symplectic group for N-dimensional IHO is usually stated to be Sp(2N, R). The important point to note is that this group is a dynamical non-invariance group for the generalized coupled system of N 1D oscillators. The example of 4D IHO system considered in \cite{8} illustrates this point: the group Sp(8, R) accounts for the dynamical symmetry for the whole energy spectrum $E<0$ for a pair of 2D IHO with attractive potential, $E>0$ for a pair of 2D IHO with repulsive potential, and $E=0$ for a pair of 2D free particle systems. 

In the case of a 2D IHO with usual attractive potential the symplectic group Sp(4, R) is not fully realized. Dirac two-oscillator system is a generalized coupled pair of 1D oscillators for which Sp(4, R)  could be envisaged. Since Sp(4, R) is a covering group of SO(3,2) the Dirac representation for SO(3,2) also becomes relevant for this case. Obviously for the Landau problem and 2D IHO, the groups Sp(4, R) and SO(3,2) are not applicable.

{\bf B4: Remarks on "the remarkable dynamical symmetry of the Landau problem"}

The dynamical symmetry of the Landau problem has been recently related with Dirac's remarkable representation of the SO(3,2) group by Dereli et al \cite{9}. The main arguments of the authors are based on (i) the identification of the generators $m_{ab}$ in \cite{10} in terms of the polynomials of the creation/annihilation operators $(a^\pm. b^\pm)$, (ii) representation of 2-component Weyl spinors in terms of these operators, and (iii) using the Majorana condition to get Majorana spinor from Dirac spinor. In addition to these, the application of Kustaanheimo-Stiefel transformation with a constraint leading to 3D Hydrogen atom - 4D IHO correspondence \cite{8}, and making use of the Jordan algebra of real matrices constitute technical tools in their work.

Regarding points (ii) and (iii) it has to be realized that already Dirac himself obtained Majorana equation \cite{16,17} in his remarkable representation of the SO(3,2) group \cite{10}. For zero mass one obtains wave equations for real Weyl spinors from Majorana equation. Therefore, most of the results obtained in \cite{9} are expected to follow from the considerations given by the authors. The crucial issue is that of linking them with the symmetry of the Landau problem.

It seems the questionable assumption is contained in point (i). The construction of the Dirac generators $m_{ab}$ from the energy and magnetic translation operators $a^\pm,~b^\pm$ respectively is no more than a formal artefact in the light of our detailed discussion on Dirac's two-oscillator system. Baskal et al \cite{11} have rightly pointed out that for a pair of operators $(a^\pm. b^\pm)$ there are only six natural generators, and four additional generators to get Dirac's ten generators one needs a coupling between the two oscillators. In the Equations (10) and (12) in \cite{9} the nature of coupling, if any, is not identified and it is not obvious. On the other hand, we have shown in the Section IV that the Landau problem reduces to 1D oscillator problem irrespective of the choice of the gauge, therefore, the claimed remarkable dynamical symmetry for the Landau problem is invalid.

\section{ \bf Discussion and Conclusion}

The remarkable representation of the 3+2 de Sitter group proposed by Dirac \cite{10} has been of great interest in building the models of elementary particles as extended structures, and the method of group contraction in connection with SO(3,2) has also been discussed in the literature. Dirac's idea in his second paper on Majorana equation \cite{17} is to overcome the unphysical implication of the Majorana equation that the spin increases to infinity as mass tends to zero, and the largest mass has zero spin. Restricting to single mass value it is found that the particle motion is some kind of a pulsating spherical shell, and the coordinates of the center of the shell are non-commuting. Discussion on orbital angular momentum and spin using gauge invariant coordinates shows that spin is zero for all momentum values of the particle. In hadron models relativistic rotator and relativistic oscillator have been studied using the group SO(3,2). We refer to one of the many papers on this subject \cite{18} that brings out the role of the group contraction and Dirac representation of the group SO(3,2) in this connection nicely.  An important result in this paper is the demonstration that the quantum relativistic oscillator can be represented using SO(3,2) spectrum generator, and the group contraction leads to a nonrelativistic 3D IHO. 

In the context of the Landau problem from \cite{16,17,18} it would seem that the group SO(3,2) or its contraction are not useful for understanding the symmetry group of the Landau problem. However, the quantum nature of the cyclotron orbit implies that the coordinates of the center of the orbit are noncommuting and the size of the orbit has a quantum limitation \cite{5}. If we visualize physical conditions where the internal structure of the Landau level and cyclotron orbit could be treated as composite of electron and magnetic field similar to composite particles postulated to explain fractional quantum Hall effect \cite{19} then it may be speculated that the SO(3,2) group and Dirac's representation \cite{10}  in the light of \cite{16,17,18} and some of the ideas using Jordan algebra and Weyl spinors \cite{9} could serve the purpose of getting new insights from the physics of Landau problem. Strong magnetic fields and many electron system having Coulomb electrostatic repulsion may also indicate the realization of two 1D oscillators having repulsive potential to complete the Sp(4,R) picture \cite{8}.

In conclusion, we have presented a thorough discussion on classical and quantum aspects of the correspondence between 2D IHO, 2D Coulomb problem and the Landau problem; new insights have been obtained on the two-oscillator system in the remarkable representation of the group SO(3,2) due to Dirac; using Wigner's group contraction it has been shown that the dynamical symmetry group for the Landau problem is the central extension of the Euclidean group $\bar{E}(2)$ and Sp(2,R) independent of the choice of the gauge; and that the claim that the dynamical symmetry group for the Landau problem is SO(3,2) is invalid. Speculative suggestion is also made regarding the possible applicability of the group SO(3,2) and Dirac two-oscillator system to the composite of electron and strong magnetic field where cyclotron orbit shows quantum nature.
 
{\bf Appendix}

In the matrix representation of the Hamiltonian (52) Dulock and McIntosh \cite{4} consider the monomials $(x, y, p_x, p_y)$ and calculate the eigenvalues and eigenvectors. The eigenvectors are obtained to be
\begin{equation}
u = {[(\omega_L^2+\omega^2) m]}^\frac{1}{2} (x+iy) +i m^{-\frac{1}{2}} (p_x +i p_y)
\end{equation}
\begin{equation}
v =  {[(\omega_L^2+\omega^2) m]}^\frac{1}{2} (x-iy) +i m^{-\frac{1}{2}} (p_x -i p_y)
\end{equation}
Linear combinations of the products of these quantities constitute the constants of motion defined as
\begin{equation}
K=\frac{(u^R v^* + u^{*R} v)}{\sqrt{R} {(u u^*)}^{\frac{(R-1)}{2}}}
\end{equation}
\begin{equation}
L=i \frac{(u^R v^* - u^{*R} v)}{\sqrt{R} {(u u^*)}^{\frac{(R-1)}{2}}}
\end{equation}
\begin{equation}
D=\frac{(u u^* - R v v^*)}{R}
\end{equation}
These constants of motion together with the Hamiltonian $H_Z$ satisfy SU(2) Lie algebra.

In the limiting case of $\omega \rightarrow 0$, the expressions (62) and (63) can be obtained from the general expressions (80) and (81) respecively. For convenience the authors denote the constants of motion as
\begin{equation}
S =( \frac{m^{1}{2}}{4i}) (v - v^*)
\end{equation}
\begin{equation}
Q= ( \frac{m^{1}{2}}{4}) (v + v^*)
\end{equation}
In the present paper, $(S,~Q)$ have correspondence with Eqs. (37) -(38).


\begin{thebibliography}{99}
\bibitem{1} H. Goldstein, Classical Mechanics (Addison-Wesley, 1980) Second Edition
\bibitem{2} L. Schiff, Quantum Mechanics (McGraw Hill, 1968) Third Edition
\bibitem{3} J. M. Jauch and E. L. Hill, On the problem of degeneracy in quantum mechanics, Phys. Rev. 57, 641 (1940)
\bibitem{4} V. A. Dulock and H. V. McIntosh, Degeneracy of cyclotron motion, J. Math. Phys. 7, 1401 (1966)
\bibitem{5} M. H. Johnson and B. A. Lippmann, Motion in a constant magnetic field, Phys. Rev. 76, 828 (1949)
\bibitem{6} E. Inonu and E. P. Wigner, On the contraction of groups and their representations, Proc. Natl. Acad. Sci. (USA) 39, 510 (1953)
\bibitem{7} U. Niederer, The maximal kinematical group of the harmonic oscillator, Helvetica Physica Acta, 46, 191 (1973)
\bibitem{8} M. R. Kibler, On the use of the group SO(4,2) in atomic and molecular physics, Molecular Physics, 102, 1221 (2004)
\bibitem{9} T. Dereli, P. Nounahon and T. Popov, A remarkable dynamical symmetry of the Landau problem, J. Phys. Conference Series, 2191, 012009 (2022)
\bibitem{10} P. A. M. Dirac, A remarkable representation of the 3+2 de Sitter group, J. Math. Phys. 4, 901 (1963) 
\bibitem{11} S. Baskal, Y. S. Kim and M. E. Noz, Einstein's $E=mc^2$ derivable from Heisenberg's uncertainty relations, arXiv: 1911.03818v1 [quant-ph] (2019)
\bibitem{12} H. V. McIntosh, On accidental degeneracy in classical and quantum mechanics, Am. J. Phys. 27, 620 (1959)
\bibitem{13} S. C. Tiwari, Pancharatnam phase for photon, Optik, 98, 32 (1993)
\bibitem{14} S. C. Tiwari, Coulomb-quantum oscillator correspondence in two dimension, pure gauge field and half-quantized vortex, Mod. Phys. Lett. A 34, 1950128 (2019)
\bibitem{15} J. B. Ehrman, On the unitary irreducible representations of the universal covering group of the 3+2 de Sitter group, Proc. Cambridge Philos. Soc. 53, 290 (1957)
\bibitem{16} P. A. M. Dirac, A positive-energy relativistic wave equation, Proc. Roy. Soc. A 322, 435(1971)
\bibitem{17} P. A. M. Dirac, A positive-energy relativistic wave equation II, Proc. Roy. Soc. A 328, 1 (1972)
\bibitem{18} H. L. Stormer, The fractional quantum Hall effect, Nobel Lecture,  Rev. Mod. Phys. 71,875 (1999)
\bibitem{19} A. Bohm, M. Loewe, P. Magnollay, M. Tarlini, R. R. Aldinger, L. C. Biedenharn and H. van Dam, Quantum relativistic oscillator. III Contraction between the algebra of SO(3,2) and the three-dimensional harmonic oscillator, Phys. Rev. D 32, 2828 (1985)
\end{thebibliography}
\end{document}